\documentclass[12pt]{article}
\title{
Glueball masses and Pomeron trajectory in nonperturbative QCD
approach}
\author{ A.B.Kaidalov and Yu.A.Simonov\\ State Research
Center\\Institute of Theoretical and Experimental Physics, \\ Moscow,
Russia}
 \date{}
  \newcommand{\be}{\begin{equation}}
 \newcommand{\ee}{\end{equation}}  
\def\fun#1#2{\lower3.6pt\vbox{\baselineskip0pt\lineskip.9pt
\ialign{$\mathsurround=0pt#1\hfil
##\hfil$\crcr#2\crcr\sim\crcr}}}

\newcommand{\ver}{\mbox{\boldmath${\rm r}$}}

\newcommand{\vep}{\mbox{\boldmath${\rm p}$}}

\newcommand{\veL}{\mbox{\boldmath${\rm L}$}}
\newcommand{\veS}{\mbox{\boldmath${\rm S}$}}
\newcommand{\veB}{\mbox{\boldmath${\rm B}$}}
\newcommand{\veE}{\mbox{\boldmath${\rm E}$}}
\newcommand{\ven}{\mbox{\boldmath${\rm n}$}}

\begin{document}
\maketitle

 \begin{abstract}
 Using a nonperturbative method based on asymptotic behaviour of
 Wilson loops  we calculate masses of glueballs and corresponding
 Regge--trajectories.
 The method contains no fitting parameters and the mass scale is
 fixed by the meson Regge slope.
   Theoretical predictions for lowest glueball states are in
  excellent agreement with lattice results. The leading glueball
 trajectory and its relation to the Pomeron is discussed in details.
 Important role of mixing between glueball and $q\bar q$ trajectories
 is emphasized.  \end{abstract}

 Glueballs are among the most  intriguing objects both in experiment
 [1]  and lattice data [2,3]. While experimental situation is not yet
 settled, lattice simulations yield an overall consistent picture
 of lowest ($<4 GeV$) mass spectrum.
 The mass scale and level ordering of the resulting glueball
 spectrastrongly differ from those of meson spectra, yielding a
 unique information about the nonperturbative (NP) structure of the
 gluonic vacuum.  Another important feature of glueball spectrum is
 that the leading glueball Regge trajectory is closely connected with
 the Pomeron singularity, which determines asymptotic behaviour of
 high--energy diffractive processes -- the object of active studies
 during last decade.

   In this paper we will study the problem of spectra of
 glueballs and of the Pomeron singularity using the method of
 Wilson-loop path integrals developed in papers
 [4,5,6]. The
 method is based on the assumption of an area law for Wilson loops at
 large distances in QCD, which is equivalent to the condition of
 confinement of quarks and gluons.
 It is shown below that long distance dynamics is defined by the
 string tension only and yields spin--averaged  glueball masses,
 while spin  splittings are due to mostly perturbative exchanges and
 relatively small.
 Our predictions for masses of lowest spin-averaged
 glueball states  in units of  ${\sqrt\sigma}$ are in an excellent
  agreement with results of recent lattice calculations [2,3] e.g.
 the calculated mass ratios $M(G)/M(0^{-+})$ are within 5\% on
 average from the lattice data of [2].  Spin-orbit and spin-spin
 splittings are calculated and found in a good agreement with lattice
 data also (within 1$\sigma$ deviation on average).

To find the Pomeron singularity,
we
calculate the leading glueball Regge-trajectory in the positive $t$
region and  extrapolate it to the scattering region of $t\le 0$.
The importance of mixing among this trajectory and $q\bar
q$--trajectories ($f,f'$) is emphasized. A calculation of these
 mixing effects leads to the  Pomeron trajectory with
 $\alpha_P(0)>1$ $(\alpha_P(0)=1.1\div 1.2)$ in accord with
 experimental observations [7]. An interesting interplay of 3 vacuum
 trajectories in the region $t>0$ is observed.

 A previous study of glueball spectrum in a similar method was done
 in [8].  In this letter we improve  results of [8] in several ways.
 First, the analysis of perturbative gluon exchanges (PGE) shows that
 PGE do not sum up to the adjoint Coulomb interaction, but rather to
 the BFKL ladder, where loop correction strongly reduces the sum, so
 that PGE can  be disregarded in the first approximation.
 Secondly, we use the rotating string Hamiltonian of [5,6] which
 yields correct  string slope $1/2\pi\sigma$ (as compared to
 $1/8\sigma$ in [8]). Finally the mixing of glueball and $f,f'$
 trajectories is taken into account, which drastically changes the
 resulting Pomeron.

 Following ref.[4], we separate gluonic fields $A_\mu$ into
 nonperturbative background $B_\mu$ and perturbative gluons $a_\mu$,
 $A_\mu=B_\mu+a_\mu$  and consider two-gluon glueballs, described by the
 Green's functions
 \be
 G_{\mu\nu, \mu'\nu'}(x,y|x',y')= \langle \Gamma^{in}
  G_{\mu\mu'}(x,
 x') G_{\nu\nu'}(y,y')\Gamma^{out}\rangle +{\rm perm}
  \ee
   where
 $\Gamma^{(in, out)}$ are operators projecting given
 quantum numbers and $G_{\mu\mu'}$ is the gluon Green's function of
 field $a_\mu$ in the background field $B_\mu$, namely
 \be
 G_{\mu\nu}(x,y) = \langle x|(-\hat D^2\delta_{\mu\nu} -2ig \hat
 F_{\mu\nu})^{-1}|y\rangle \ee where $\hat D_{\mu}=\partial_\mu-ig
 \hat B_\mu, ~\hat F_{\mu\nu}$ is the field strength of the field
 $\hat B_\mu$ in the adjoint representation, and averaging over
 background $B_\mu$ is implied by angular brackets.

 Expression (1) can be written as the path integral (see refs.
 [4,8])
 \be
 G_{\mu\nu,\mu'\nu'}(x,y|x',y')=const \int^\infty_0 ds \int^\infty_0
 ds' DzDz'e^{-K-K'}\langle W_F\rangle
 \ee
 where $K=\frac14\int^s_0(\frac{dz}{d\tau})^2 d\tau,~~K'$ is the same
 with primed $z,\tau,s$, and
 \be
 \langle W_F\rangle =tr P_BP_F\langle \exp \{ig \int_C B_\mu du_\mu +2ig
 \int^s_0\hat F d\tau +2ig \int^{s'}_0\hat Fd\tau'\}\rangle
 \ee
  Here
 $P_B,P_F$ are ordering operators of the color matrices $B_\mu$ and
 $\hat F$ respectively.  The terms with
 $\hat F$ generate spin--dependent nonperturbative contributions,
which are calculable and relatively small, and we shall treat those
 terms perturbatively.

 Neglecting $\hat F's $ as a first approximation, one arrives at the
 Wilson loop in the adjoint representation, for which one can use the
 minimal area law, confirmed by numerous lattice data [9] up
 to the distance of the order of $1 fm$,
  \be
  \langle W_{adj}\rangle
 =Z~ exp (-\sigma_{adj}S_{min})
 \ee
  where we have included in $Z$
 self-energy and nonasymptotic corrections, since (5) is valid for
 large loops with size $r, T\gg T_g$, where $T_g$ is the gluon
 correlation length.

 Applying now the general method of
 [5] to the Green's function (3),
 one defines the Hamiltonian $H_0$ for 2 gluons without spin
 corrections.
 $$
H_0=\frac{p_r^2}{\mu(t)}+
\mu(t)+\frac{L(L+1)}{r^2[\mu(t)+2\int^1_0(\beta-\frac12
)^2\nu(\beta, t) d\beta]}+
$$
\be
+\int^1_0\frac{\sigma^2_{adj}d\beta}{2\nu(\beta,t)} r^2 +\frac12
\int^1_0 \nu(\beta, t) d\beta
\ee
Here $L$  is the angular momentum in the c.m. system of two gluons,
${\bf L}=[{\bf r}\times {\bf p}], ~~ r$ is the distance between
gluons, and ${\bf p}, p_r$ are momentum and its radial projection
respectively; $t$  and $\beta$ are the c.m. time and distance along
the string -- two coordinates on the world sheet of the string.

 Positive auxiliary
functions
 $\mu(t)$ and $\nu(\beta,t)$
 are to be found from the  extremum condition
[5].  Their extremal values are equal to the  effective gluon
energy $\langle \mu\rangle $ and energy density of the adjoint string
 $\langle \nu\rangle $.The  physical meaning of those is
 better illustrated by limiting cases of the free gluon, where
 $\langle\mu\rangle=\langle \sqrt{{\bf p}^2}\rangle$ and of the free
 string without  quarks, where
 $\langle\nu\rangle= \sqrt{\frac{8\sigma
 L}{\pi}}(1-4(\beta-\frac12)^2)^{-1/2}$.

For the case  $L=0$ the extremization over $\mu$ and $\nu$ yields a
simple answer, coinciding with the Hamiltonian of the relativistic
potential model
\be
H_0=2\sqrt{p^2_r} + \sigma_{adj} r
\ee
With  the replacement of the
operators $\mu(t), \nu(\beta,t )$ (which by extremization are
expressed through
operators $p, r$) by $c$ --numbers, to be found
from extremization of eigenvalues of $H_0$,
one obtains a simple form,
used in [8],
\be
H'_0= \frac{\vep^2}{\mu_0} +\mu_0+\sigma _{adj}r
\ee
The  eigenvalues of (8)
are about 5\% higher than those of $H_0$.

The value of $\sigma_{adj}$ in (8) can be found from the string
tension
$\sigma_{fund}$ of $q\bar q$ system,
multiplying it  by $\frac{9}{4}$, as it follows from Casimir scaling
observed on the lattice [9]. Taking  experimental Regge slope for
mesons $\alpha'=0.89~ GeV^2$ one obtains $\sigma_{fund}=0.18~ GeV^2$
and $\sigma_{adj}\approx 0.40~ GeV^2$.

In what follows the parameter $\mu$ and its optimal value $\mu_0$,
which is computed from eigenvalues of (8) and proportional to
$\sqrt{\sigma_{adj}}$, play very important role.  The
way they enter spin corrections (see below) and magnetic moments
shows that $\mu_0$ is an effective (constituent)
gluon mass (or constituent quark mass in the equation for the $q\bar
q$ system).

Note that our lowest "constituent gluon mass" $\mu_0(n=L=0)=0.528
~GeV$ (for $\sigma_f=0.18~ GeV)$ is
not far from the values
introduced in the potential models, the drastic difference is that
our $\mu_0$ is calculable and  depends on $n,L$ and grows for higher
states.

The mass spectrum   of
$H_0$, Eq.  (6), and was studied in [5,11]. With a good
accuracy it can be described by a very simple formula
\be
\frac{M^2}{2\pi\sigma_{adj}}=L+2n_r+c_1
\ee
where $L$ is the orbital momentum, $n_r$ --radial quantum number
and $c_1$ is a constant $\approx 1.55$. It describes an infinite set
of linear Regge-trajectories shifted by $2n_r$ from the leading one
($n_r=0)$.
The only difference between light quarks and gluons is the value of
$\sigma$, which determines the mass scale. A similar spectrum was
obtained independently by numerical quantization of $\bar q q$
sysytem in [12].

The lowest glueball state with $L=0, n_r=0$ has $M=2.01~ GeV$.
It corresponds to a degenerate $0^{++}$ and $2^{++}$ state.

In order to compare our results with the corresponding lattice calculations
[2,3]
 it is convenient to consider the quantity
$\bar M/\sqrt{\sigma_f}$, which is not sensitive to the choice of
string tension $\sigma$
 \footnote{Note that the values
$\sigma_f\simeq 0.21-0.23~ GeV^2$ used in lattice calculations differs
by about 20\% from the "experimental" value $\sigma_f=0.18~
GeV^2$.}.
 We also introduce the spin averaged mass $\bar M$, which for
$L=0, n_r=0$ states is defined as
$\bar M=\frac13 (M(0^{++})+2M(2^{++}))$, and in a similar way for
higher states. This definition takes into account the structure of
spin-splitting terms (see below in (10)), so that $\bar M$ can be
compared with the eigenvalues $M_0$ of spinless Hamiltonian (6).

The comparison of our predictions for spin averaged masses of the
lowest glueball states with corresponding lattice results is given in
Table 1.  For
the average mass with  $L=2, n_r=0$  lattice
results are limited to the state  $3^{++}$.

It follows from Table 1 that the spin--averaged masses
obtained from purely confining force with relativistic kinematics for
valence gluons are in a good correspondence with lattice data, which
implies that
PGE effects are suppressed. We come back to this point in what
follows.

    Now we shall consider spin splittings of glueball masses
   and  shall treat spin effects as a small perturbation.
    Such a
   treatment is justified aposteriori  by our results and by lattice
   data, which demonstrate that spin splittings for glueball states
   (apart from $2^{++}-0^{++}$) amount to less than 10-15\% of the
   total mass.

   To proceed one should choose among two possibilities of spin state
   description of bound gluons. One can insist on transversality
   condition ( valid for a free gluon) also for gluons in the
   glueball, as  it is  done in the potential model of [10].
    Instead in our  formalism the bound gluon becomes
   massive due to the string ( with the effective energy $\mu_0$  not
   equal to the gluon momentum $|{\bf k}_1|)$ and has 3 spin
   polarizations, similarly to $W^{\pm}, Z^0$ bosons, which get their
   mass from Higgs condensate. For more discussion see [11].

   Then two--gluon mass operator can be written as
   \be
   M=M_0(n,L) +\veS\veL M_{SL}+\veS^{(1)}\veS^{(2)} M_{SS}+M_T,
   \label{7.1}
   \ee
   where $ \veS=\veS^{(1)}+\veS^{(2)}$,  $M_0$ is the eigenvalue of
   the Hamiltonian $H\equiv H_0+\Delta H_{pert}$, and $H_0$ is given
   in (7) (or its approximation in (8)), while $\Delta H_{pert}$ is
   due to PGE.

   To obtain three other terms in (10) one should consider averaging
   of the operators $\hat F$ in the exponent of (4) and take into
   account that
   \be
   -2i\hat F_{\mu\nu}=2(\veS^{(1)} \veB^{(1)}+ \tilde{\veS}^{(1)}
   \veE^{(1)})_{\mu\nu}
   \label{7.2}
   \ee
   and similarly for the term in the integral $\int\hat F d\tau'$,
   with the replacement of indices $1\to 2$.
Here gluon spin operators are introduced, e.g.
\be
(S^{(1)}_m)_{ik}=-ie_{mik},~~i,k =1,2,3, (\tilde S^{(1)}_m)_{i4}
=-i\delta_{im}
\label{7.3}
\ee

Now the spin splittings are obtained from (4) where the proper time
$(\tau,\tau'$) is replaced by the real (Euclidean) time $(t,t')$ via
the relation [4-6] $d\tau=\frac{dt}{2\mu(t)}$, where $\mu(t)$ is the
same operator as in the Hamiltonian (6) with its extremal value
$\mu_0$ playing the role of the gluon constituent mass ( and
explicitly calculated from (6) or (8)).
As a result the spin--dependent
part of the Hamiltonian(10) can be written  in the form similar to
that of Eichten and Feinberg [12]
(see [11] for more details)
 $$ \Delta H_s=
\frac{\veS\veL}{\mu^2_0} (\frac1r \frac{dV_1}{dr}
+\frac1r\frac{dV_2}{dr})+ \frac{\veS^{(1)}\veS^{(2)}}{3\mu^2_0} V_4
(r)+
$$
\be + \frac{1}{3\mu_0^2} (3(\veS^{(1)}\ven)(\veS^{(2)}\ven)
   -\veS^{(1)}\veS^{(2)})V_3(r)+\Delta V
   \label{7.4}
   \ee
   where $\veS=\veS^{(1)}+\veS^{(2)},\Delta V$ contains higher
   cumulant contributions which can be estimated  of the
   order of 10\% of the main term in (4) [4].

   The functions $V_i(r)$ are the same as   for heavy quarkonia [13]
   and  the NP part of spin splittings can be in this way expressed
   through the field correlators $D(x), D_1(x)$ [14], measured on the
   lattice [15]. The resulting estimates show that the only
   appreciable  NP spin--splitting is due to $V_1, V_2$ (Thomas
   precession) and was computed numerically as in [13].

   Perturbative contributions to spin splittings can be calculated in the
same way using
results of refs [10]. Combining all corrections and values of
$M_0$ from Table 1 we obtain glueball masses compared with lattice
data in Table 2 for $\sigma_f=0.23~ GeV^2$.

   The general feature of spin--dependent contribution $\Delta H_s$
   is that it dies out  with the growing orbital or radial
   number. This feature is well supported by the lattice data in
   Table 2.

   One can see in Table 2 that magnitude of our spin--spin and
   spin--orbit splittings ( proportional to
   $\langle\delta^{(3)}(r)\rangle$ and  $\langle 1/r^3\rangle $
   respectively) is in good agreement with lattice data
   (our splittings are on average some 30\% higher than lattice spin
   splittings defined with rather large uncertainties); inclusion of
   PGE in the form of adjoint Coulomb would increase splittings
   several times [8] and destroy the agreement, yielding another
   phenomenological argument against adjoint Coulomb form of PGE.

In many analytic calculations of glueball masses it is postulated that there
is a Coulomb-type interaction between valence gluons, which differs from the
$q\bar q$ case by
the same $\frac94$ factor as in eq. (9).
If this would be true it would
lead to a drastic decrease of glueball masses calculated above due to
 Coulomb attraction, especially for L=0 (for
$\alpha_s=0.3$ this mass drops down by 0.5 GeV [8]).   This clearly
contradicts lattice calculations and casts a  doubt on the validity
of the assumption about  presence of  Coulomb
interaction (without higher loop corrections).

A careful study of general derivation of Coulomb-like small distance
 contribution in this formalism shows [11]
  that contrary to
 the quark case for valence gluons  perturbative
gluon exchanges do not  sum up to the Coulomb potential, but rather
to the BFKL ladder [16].

 In order to estimate effects of small distance contributions we shall use
 the analysis of these effects on gluonic Regge--trajectories not from the
 glueballs mass spectra at positive $t$, but for $t=0$. Extensive
 calculations of the gluonic Pomeron trajectory intercept have been
 carried out in the leading log approximation (LLA)
 [16]
and $\alpha_s$ cocorrections were calculated recently
[17].

An intercept of the leading Regge pole is strongly modified by
$\alpha_s$ corrections
$$
\Delta=\alpha_P(0)-1=\alpha_s\frac{12}{\pi}ln 2(1-C\alpha_s)
$$
The coefficient $C$ is large ($\approx 6.5$) and $\alpha_s$
correction substantially decreases $\Delta$ compared to  LLA
result.

The coefficient $C$ is rather large $(\approx 6.5)$ and the $\alpha_s$
correction strongly reduces $\Delta$. Its value depends on the renormalization
scheme and scale for $\alpha_s$. In the "physical" (BLM) scheme values of
 $\Delta$ are in the region $0.15\div 0.17$ [18].
 We
 can estimate mass-shift of the lowest glueball state
 due to PGE effects using this
 result and assuming that the slope $\alpha'_P\approx 0.4~ GeV^2$ of
eq.(11) is not be strongly modified by perturbative effects.
Thus one can expect that characteristic shift due to perturbative
effects in $\bar M^2(L=0,n_r=0),\delta\bar M^2\approx
\Delta/\alpha'_{adj}\approx(0.38\div 0.48)~GeV^2$. This corresponds to the
shift in $\bar M(L=0,n_r=0),\delta\bar M\approx \delta\bar M^2/2\bar
M\approx 0.1~ GeV$. This shift is much smaller than naive estimate
using adjoint Coulomb interaction and gives some justification to the
neglect of PGE in our calculations. It should be noted that this is
 only a rough estimate of the perturbative effects.

 The three--gluon system may be considered in the same way, as it was
 done for the two-gluon glueballs. The $3g$ Green's function
 $G^{(3g)}$ is obtained as the background--averaged product of 3
 one--gluon Green's function, in full analogy with (1). Assuming large
 $N_c$ limit for simplicity and neglecting spin splittings one
 arrives at the path integral  (cf equation (3))
 \be
 G^{(3g)}=const \prod^3_{i=1}\int^\infty_0 ds_i  Dz^{(i)}
 e^{-K_i-\sigma S_i}
 \ee
 where $\sigma \equiv \sigma_{fund}
 $, since every gluon is connected by a fundamental string with each
 of its neighbours.

 Using as before  method of refs. [4-6]
  one obtains the following Hamiltonian
 (we assume symmetric solution with equal $\mu_i(\tau)\equiv
 \mu(\tau), i=1,2,3$ (no orbital excitations was assumed as in (8))
 \be
 H^{(3g)}=\frac{\vep^2_\eta+\vep^2_\xi}{2\mu}+\frac{3\mu}{2}+
 \sigma(|\ver_1-\ver_2|+|\ver_2-\ver_3|)
 \ee
 Here $\vep_\eta,\vep_\xi$ are Jacobi momenta
 for the 3--body system of ($q,\bar q, g)$.

To simplify treatment further, we shall consider $\mu$ as a constant to
be found from the extremum of eigenvalues, as in (8), which in that
case provided some 5\% increase in eigenvalues and  we expect the
same situation in this case.

The eigenvalues of $H^{(3g)}$ were found using the hyperspherical
method introduced in [19].  A reliable estimate of $M$ gives for
the lowest state (and $\sigma_f=0.18~ GeV^2 $) the value $M_0=3.22~ GeV$
(3.62 for $\sigma=0.23~ GeV^2$).  Perturbative hyperfine interaction
  increases the mass of the $3^{--}$ spin splittings compared to
 this value by $\approx 0.29~ GeV$. The large value for the lowest
 three-gluon state is in an agreement with lattice calculations (see
 Table 2). For more details concerning $3g$ states see [11].

Let us consider now the problem of the Pomeron Regge trajectory
 in more details, taking into account both  nonperturbative and perturbative
contributions to the Pomeron dynamics.

The large distance, nonperturbative contribution gives according to Eq.(12)
for the leading glueball trajectory ($n_r=0$ )
\be
\alpha_P(t)=-c_1+\alpha'_P t +2
\ee
with $\alpha'_P=\frac{1}{2\pi\sigma_a}$.

In eq.(16) we took into  account spins of "constituent"
gluons, but neglected  a small nonperturbative spin-spin
interactions.
For an intercept of this trajectory we obtain
$\alpha_P(0)\approx 0.5$, which is substantially below the value
found from analysis of high-energy interactions
[7]
$\alpha_p(0)=1.1\div 1.2$.

The perturbative (BFKL) contribution leads to a shift (increase) of the Pomeron
intercept by $\approx 0.2$ as it was explained above. The resulting
$\alpha^{(0)}_P(0)\approx 0.7$ is still far from experiment.

 The most important
nonperturbative source, which can lead to an increase of the
Pomeron intercept
 is, in our opinion, the
quark-gluon mixing or account of quark-loops in the gluon "medium".
In the 1/N -expansion the effect is proportional to $N_f/N_c$, where
$N_f$ is the number of light flavours

In the leading approximation of the $1/N_c$-expansion there are 3
Regge-trajectories with vacuum quantum numbers,
- $q\bar q$-planar trajectories ($\alpha_f$ made of
$u\bar u$ and $d\bar d$ quarks, $\alpha_{f'}$ made of $s\bar s$-quarks) and
pure gluonic trajectory - $\alpha_G$. The transitions between quarks and
gluons $\sim\frac{1}{N_c}$ will lead to a mixing of these trajectories.
Note that for a realistic case of $G,f$ and $f'$-trajectories (Fig.1)
all 3-trajectories before mixing are close to each other in the small
$t$ region.  Trajectory of gluonium crosses planar $f$ and
$f'$-trajectories in the positive $ t$ region $(t<1~ GeV^2)$. In this
region mixing between trajectories is essential even for small
coupling matrix $g_{ik}(t).$
Lacking calculation of these effects in QCD we will consider them in
a semi--phenomenological manner.

 Denoting by $\bar \alpha_i$ the bare  $f,f'$ and $G$--trajectories
 and introducing the mixing matrix $g_{ik}(t)(i,k=1,2,3)$ we obtain
 the following equation for determination of resulting trajectories
 after mixing [11]
  $$ j^3-j^2\sum\bar \alpha_i+j(\sum_{i\neq k}\bar
 \alpha_i\bar \alpha_k-g^2_{ik})-
 \bar\alpha_1\bar\alpha_2\bar\alpha_3+
 \sum_{i\neq k\neq l}\bar \alpha_lg^2_{ik}-2g_{12}g_{13}g_{23}=0
 $$
  For realistic values of $g_{ik}(t)$ (for details see [11]) the
  resulting trajectories are shown in Fig. 1 by solid lines. The
  pomeron intercept is shifted to the values $\alpha_P(0)>1$. For
  $t>1 GeV^2$ the Pomeron trajectory  is very close to the planar
  $f$--trajectory, while the second and third vacuum trajectories --
  to $\alpha_{f'}$ and $\alpha_G$ correspondingly.

 Let us consider now the "odderon"--Regge trajectory --
the leading gluonic  trajectory with negative signature and
$C$-parity.  Mass of the lowest $3g$ glueball with spin $3^{--}$
corresponding to this trajectory has been estimated
above and found to be large $\approx 3.6~ GeV$ in accord with
lattice data. The slope $\alpha'_{3g}$ for this trajectory should be
equal to the one of $gg$--trajectory
 \footnote[3]{The situation is
analogous to the case of $q\bar q$(meson) and $qqq$ (baryon) Regge
trajectories, since both for baryons and $3g$ glueballs at large $L$
the structure $q-2q(g-2g)$ is energetically preferrable. } and thus
 the intercept of the nonperturbative glueball "odderon" is very low
$\alpha_{3g}(0)\approx -1.6$.  Mixing with $q\bar q$--trajectories
$(\omega,\varphi)$ is much smaller than in the Pomeron case as there
is no crossing of the odderon  and $(\omega,\varphi)$ trajectories in
the small t-region.

The main results of the paper can be summarized as follows.
We calculated the 2g and 3g glueball spectrum analytically and compared
resulting masses with lattice data, finding an excellent   agreement,
within 5\% on average for ratios of glueball masses.  We stress that
our spectrum contains no arbitrary fitting parameters, since  all
 masses are expressed in terms of the string tension, which is fixed
 by the experimental meson Regge slope.  A smallness of Coulomb-type
 contributions to glueball masses is pointed out.  The spin
 splittings of glueball masses were obtained from first perturbative
 corrections calculated with nonperturbative wave functions. There is
a good agreement on spin splittings between our calculations and
lattice data as well.  This indicates that the main ingredient of the
glueball dynamics is the adjoint string occurring between gluons in
the two-gluon glueballs.  The string dynamics implies that glueball
spectrum corresponds to straight-line Regge trajectories, which have
the Regge slope equal to $\frac{4}{9}$ of that for meson
trajectories.  For the Pomeron Regge trajectory we found that with an
account of only gluonic contributions (both nonperturbative and
perturbative) the intercept is below unity and in order to obtain a
phenomenologically exceptable value of the intercept it is necessary
to take into account mixing between gluons and $q\bar q$ pairs.  We
note that both nonperturbative (string dynamics, quark loops) and
perturbative effects are important to obtain $\alpha_P(0)>1$. In this
approach the "soft" and "hard" dynamics are strongly intermixed to
 produce the leading  Pomeron pole.

 The partial support from the RFFI grant 96-15-96740 is gratefully
             acknowledged. One of the authors (A.K.) acknowledges a
             support of the  NATO grant OUTR.LG 971390.

\newpage

\begin{center}

{\bf Table 1}\\
 Spin averaged glueball masses $M_G/\sqrt{\sigma_f}$

\vspace{1cm}

 \begin{tabular}{|l|l|l|l|l|} \hline
\multicolumn{2}{|c|}{ Quantum}& This& \multicolumn{2}{c|}{Lattice data}\\
\cline{4-5}
\multicolumn{2}{|c|}{ numbers}&work& ref. [3]&ref. [2]\\\hline
 &$l=0,n_r=0$&4.68&4.66$\pm$0.14&4.55$\pm$0.23\\   \cline{2-5}
 2 gluon&$l=1,n_r=0$&6.0 &6.36$\pm$0.6 &6.1 $\pm$0.5 \\\cline{2-5}
 states&$l=0,n_r=1$&7.0 &6.68$\pm$0.6 &6.45$\pm$0.5 \\\cline{2-5}
 &$l=2,n_r=0$&7.0 &9.0 $\pm$0.7(3$^{++}$) &7.7 $\pm 0.4(3^{++})$ \\
 \cline{2-5}
 &$l=1,n_r=1$&8.0 &  &8.14 $\pm 0.4(2^{*-+})$ \\ \hline
 3 gluon&K=0&7.61&&8.19$\pm$0.48\\
 state&&&&\\\hline
   \end{tabular}
\vspace{1.5cm}

{\bf Table 2}\\

Comparison of predicted glueball masses with lattice data (for
$\sigma_f=0.238 GeV^2  $

\vspace{0.5cm}

 \begin{tabular}{|l|l|l|l|l|l|r|} \hline
$J^{PC}$& M(GeV)&  \multicolumn{2}{c|}{Lattice
 data}
 &\multicolumn{2}{c|}{$M[G]/M[0^{-+}]$}
 &Difference\\ \cline{3-4}   \cline{5-6}
  &This   work& ref.  [2]& ref. [3]
  &This   work& ref.  [2]&\\\hline

$0^{++}$&1.58&1.73$\pm$0.13&1.74$\pm$0.05
&0.62&0.67(2)&-7\%\\
$0^{++*}$&2.71&2.67$\pm$0.31&3.14$\pm$0.10
&1.06&1.03(7)&3\%\\
$2^{++}$&2.59&2.40$\pm$0.15&2.47$\pm$0.08
&1.01&0.92(1)&9\%\\
$2^{++*}$&3.73&3.29$\pm$0.16&3.21$\pm$0.35
&&&\\
$0^{-+}$&2.56&2.59$\pm$0.17&2.37$\pm$0.27
&&&\\
$0^{-+*}$&3.77& 3.64$\pm$0.24&
&1.47&1.40(2)&5\%         \\
$2^{-+}$&3.03&3.1$\pm$0.18&3.37$\pm$0.31
&1.18&1.20(1)&-1\%\\
$2^{-+*}$&4.15& 3.89$\pm$0.23&
&1.62&1.50(2)&8\%         \\
$3^{++}$&3.58&3.69$\pm$0.22&4.3$\pm$0.34
&1.40&1.42(2)&-2\%\\
$1^{--}$&3.49& 3.85$\pm$0.24&
&1.36&1.49(2)&-8\%         \\
$2^{--}$&3.71& 3.93$\pm$0.23&
&1.45&1.52(2)&-1\%         \\
$3^{--}$&4.03& 4.13$\pm$0.29&
&1.57&1.59(4)&         \\\hline
  \end{tabular}
   \end{center}
\newpage
\begin{center}
{\bf Figure Caption}\\
\end{center}
\vspace{0.2cm}
{\bf Fig.1} Vacuum trajectories before mixing (dotted lines) and after mixing
 (solid lines).\\

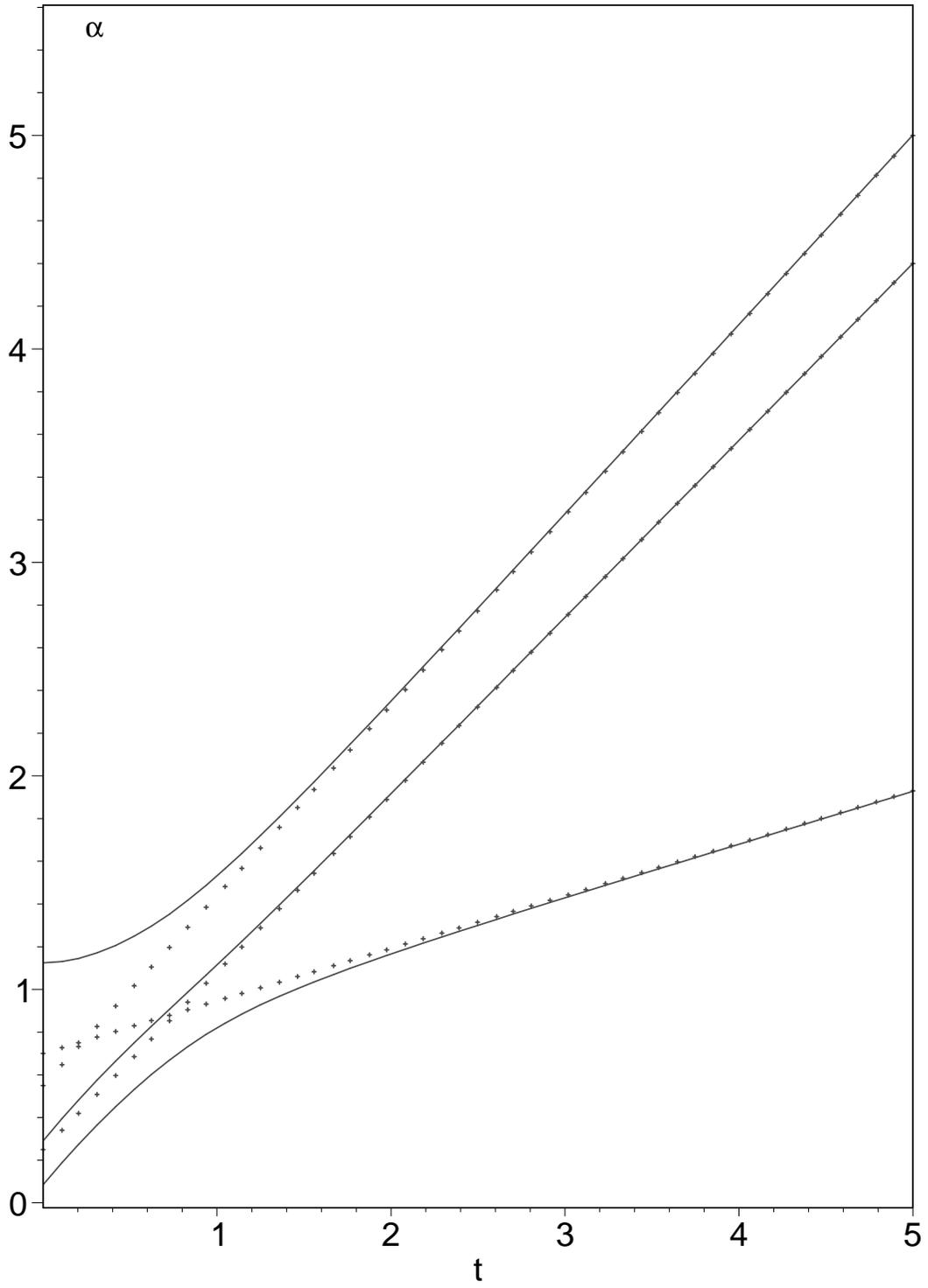
\end{document}